# EIS2MOD: a DRT-based modelling framework for Li-ion cells

Pietro Iurilli, Claudio Brivio, Rafael E. Carrillo, Vanessa Wood

*Abstract*— The correct assessment of battery states is essential to maximize battery pack performances while ensuring reliable and safe operation. This work introduces EIS2MOD, a novel modelling framework for Li-ion cells based on Distribution of Relaxation Time (DRT). A physically based Electric Circuit Model (ECM) is developed starting from Electrochemical Impedance Spectroscopy (EIS) and Open Circuit Voltage (OCV) measurements. DRT is applied to deconvolve the electrochemical phenomena from the EIS. The presented methodology is based on: i) DRT calculation from EIS, ii) DRT analysis for ECM configuration and iii) Model parameters extraction and fitting. The proposed framework is applied to large format Li-ion pouch cells, which are tested over the whole State of Charge (SoC) range and a wide temperature range (-10°C to 35°C). Different current profiles have been tested to validate the model, showing its high accuracy in reproducing the battery cell behavior (e.g. RMSE on the battery terminals voltage lower than 1.50% for driving cycle simulations at variable temperature and SoC). An additional advantage of EIS2MOD is its light computational load thus offering an attractive framework for battery management system implementation.

*Index Terms*— Battery modeling, Electrochemical Impedance Spectroscopy, Distribution of Relaxation Time, Li-ion battery

## I. INTRODUCTION

THE correct assessment of Li-ion battery cell states is essential to maximize battery pack performances while ensuring reliable and safe operation [1]. For this purpose, voltage and current signals are sensed by the Battery Management System (BMS) and combined to estimate indicators such as the State of Charge (SoC), i.e. the fraction of cell's available capacity and the State of Health (SoH), i.e. the remaining capacity with respect to original one. The practical implementations adopted in BMS can be classified in three main groups [2]–[5]: (i) *Coulomb Counting methods*, in which a simplified analytical representation of the battery (i.e. current integration) is used to compute the SoC and to update the SoH by referring to manufacturer's datasheets; (ii) *model-based estimation methods* in which battery cell models are used to infer about the SoC/SoH and online estimators or adaptive filters are used to account for measurements errors and correct deviations on the estimation; and (iii) *data-driven based methods* in which an input-output approach (i.e. black-box models) is used to develop estimators; typical examples of this approach are fuzzy controllers, neural networks and support vector machines.

The above solutions differentiates when it comes to evaluate the trade-off between accuracy and computational effort. For instance, coulomb counting methods are well suited when looking for low computational effort, but their simplicity comes at the expense of low accuracy. On the other side, data-driven based methods could give very accurate results but they require intensive computing and offline learning processes based on specific and high quality data in large quantity [6]. Between these two solutions, model-based approaches are mostly used in literature due to the interpretability of the models [6], and their capability to reflect the electrochemical characteristics of the cell [7], [8]. Among all possibilities, Electric Circuit Models (ECMs) show a good balance between accuracy and model complexity [7].

In ECMs the response of the battery at the two terminals is represented by a combination of circuit elements [9]. Model's configurations ranges from simple ones, where a voltage source and a resistor are used to simulate the Open Circuit Voltage (OCV) and total internal resistance of the cell, to more complex ones, where a chain of resistors and capacities is used to simulate the full dynamic behavior of the cell [7]. In all cases, the model's parameters must be identified based on dedicated measurements. Electrochemical Impedance Spectroscopy (EIS) is widely used for this purpose [10]–[13]. EIS is a non-invasive technique that characterize the electrical properties of different material and their interfaces by means of low amplitude sinusoidal voltage or current excitation over a range of frequencies [14]. The most common approach to derive the ECM is to choose the series connection of elements according to the shape of the impedance curve and followed by a Nonlinear Least Square parameter fitting [15]. Therefore, the model derived by EIS is based on a-priori assumption about the circuit configuration: expertise on electrochemical processes is required to determine the type and number of elements necessary to simulate the cell behavior [7], [16]. However, it is often difficult to interpret the impedance spectra: the overlapping of different processes in the same frequency range

P. Iurilli is with the PV-Center, CSEM, Neuchâtel, Switzerland and with the Department of Information Technology and Electrical Engineering, ETH Zürich, Zürich, Switzerland (pietro.iurilli@csem.ch).

C. Brivio and R. E. Carrillo are with PV-Center, CSEM, Neuchâtel, Switzerland.
V. Wood is with the Department of Information Technology and Electrical Engineering, ETH Zürich, Zürich, Switzerland.

makes difficult the choice of adequate circuit elements [17]. Moreover, the model validation with respect to experimental measurements could lead to unnecessary over-parameterization [11], [18].

A way to overcome these issues is to apply the Distribution of Relaxation Time (DRT) method. DRT is a mathematical transformation that allows to find the time constants of the physical processes occurring inside an electrochemical system [18]. The EIS spectra are deconvoluted in the time domain making possible to distinguish major and minor polarization effects that are normally overlapped in the frequency domain. These effects are represented by peaks with different magnitude and time constants. Those peaks allows to a better visualization of the individual processes occurring in the system under investigation and provide direct access to their kinetic parameters [19]. In this way, DRT representation provides higher resolution than classical impedance representations such as Nyquist or Bode plot. This technique has been widely applied in literature. For instance, Sabet et al. investigated the DRT profiles of pristine and aged Lithium Nickel Cobalt Aluminum Oxide (NCA) and Lithium Nickel Manganese Cobalt Oxide (NMC) cells, respectively in [20] and in [21]. In both the cases, the authors were able to define the main processes occurring in the cells and to track the cell degradation by analyzing the peaks variation of DRT profiles. The variation in position and magnitude of the DRT peaks has been also analyzed in [22] and [23]. In both the two cases the authors computed DRT profiles from EIS at different temperatures for new and aged cells. As in the previous case, the analysis of DRT profiles has been used to detect degradation processes occurring in the cell. However, the same results were not used to develop a model of the cell under investigation. In [15], Illig developed a very detailed physically based model of Lithium Iron Phosphate (LFP) cells exploiting DRT profiles derived by EIS and time domain measurements. The model was based on dedicated lab-scale cells to identify separately the electrochemical loss processes of the anode and of the cathode and it was used to analyze and improve the electrodes' performances. Another approach related to modelling, is presented by Schmidt et al. in [24]. The authors introduce a procedure to develop ECMs of Li-ion cells from DRT derived by combination of EIS and pulsed test. The model validation is performed through simulation in the time domain at constant temperature. The number of circuit elements is defined by the user: the higher the number of elements, the higher the accuracy but also the computational effort. Thus, there is no connection between the physical processes occurring in the cell and the circuit elements used to simulate its behavior.

This work presents EIS2MOD, an extension of the work presented in [25]. EIS2MOD is a novel modelling framework based on DRT that goes automatically from EIS and OCV measurements to a complete battery model. DRT is applied to commercial Li-ion cells (and not only to lab-scale formats as in [15]) as an insightful tool to: (i) identify the physical processes occurring inside an electrochemical cell, and (ii) to model the detected processes by means of a physically-based ECM. By doing so, each circuit element has a clear physical meaning related to different processes, differently from the solution proposed in [24]. Moreover, the modelling framework is an attractive option for BMS implementation since it offers a high model accuracy while keeping a low computational load and a simple implementation.

The work is structured as follows. Section II introduces the different steps of the novel methodology, from DRT calculation to ECM configuration, to parameters extraction and fitting. Section III details the description of the experimental tests on large format pouch cells and the model developed applying the novel methodology. The cell characterization has been performed over the whole SoC range at four different temperatures: -10°C, 5°C, 20°C and 35°C. Finally, Section IV presents the results of the model validation with different current profiles at constant and variable temperature.

## II. METHODOLOGY

The schematic representation of the methodology presented in this work is shown in Fig. 1. It is divided in two main blocks: model development and model validation. The first block is the EIS2MOD modelling framework and includes three main steps: A) DRT calculation, B) DRT analysis for ECM configuration and, C) parameters extraction and fitting. The second block, i.e. the validation process, is necessary to benchmark the model and assess its performance compared to real cells' measurements. Three current profiles have been used to perform the model validation: full discharge at constant current rate, charging and discharging steps at different current rates and driving profile simulation derived by the dynamic discharge profiles for Battery Electrical Vehicles (BEV) from the international standard on performance testing of Li-ion cells for electric vehicles use [26]. More details on the validation process and on the testing protocols will be given in Section IV. In the following we detail the main steps of the model development block.

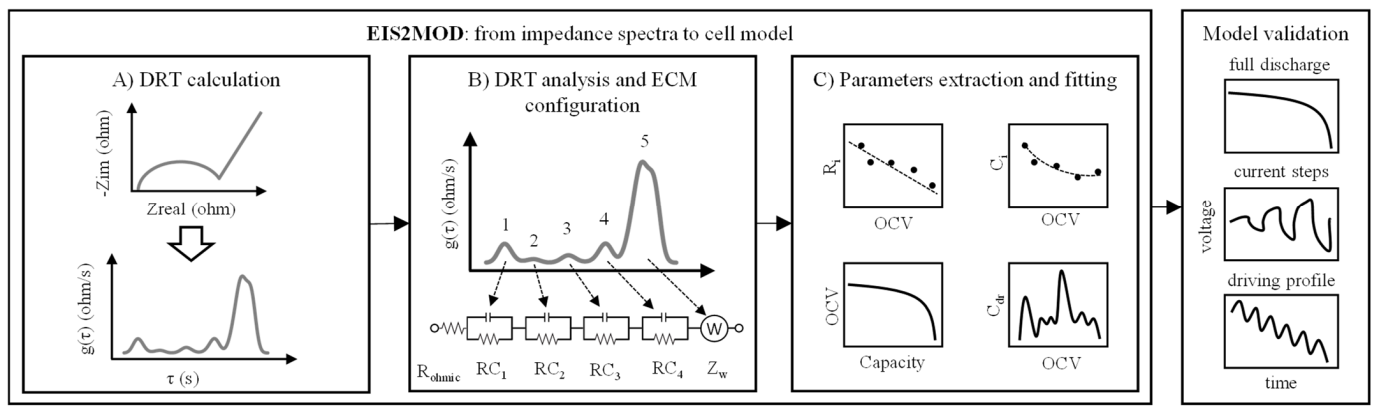

Fig. 1. Schematic representation of the methodology presented in this work. The EIS2MOD modelling framework is based on three steps: A) the calculation of DRT profile from EIS spectrum; B) the analysis of DRT profile, attributing physical processes to different peaks and making the configuration of the ECM; C) the extraction and fitting (linear or quadratic) of parameters from DRT peaks and calculation of intercalation capacitance from the OCV curve. The model validation is performed with three different profiles: full discharge at constant rate, current steps at different rate and driving profile simulation derived by IEC 62660-1 [23].

*A. DRT calculation*

The DRT method allows the deconvolution of EIS measurements into the multiple polarization processes occurring in an electrochemical system. The DRT function is normally depicted as $g(\tau)$ and is plotted against the time constants axis (as shown in Fig. 1) resulting in a profile with variable number of peaks that describe the properties of the studied system. The area below $g(\tau)$ peaks represents the polarization resistance of the system under investigation. High quality measurement and time invariance are required to correctly apply the DRT method. A suitable tool to check the validity of the impedance spectra is to apply the Kramers-Kronig criterium [27]. DRT calculation can be performed when the residuals between measured and Kramers-Kronig reconstructed spectrum are below 1% [28], [29].

Once linearity and time invariance are fulfilled, it is possible to represent the impedance as an infinite number of RC-elements in series with a resistor:

$$Z(j\omega) = R_{ohmic} + \int_0^\infty \frac{g(\tau)}{1+j\omega\tau} d\tau, \quad (1)$$

where $R_{ohmic}$ represents the ohmic resistance and $g(\tau)$ is the function that represents the time relaxation characteristics of the electrochemical system under evaluation and that satisfies the conditions of non-negativity (total and partial polarization resistances are positive for most of the electrochemical systems) [30]. The integral in (1) is often normalized, leading to:

$$Z(j\omega) = R_{ohmic} + R_{pol} \cdot \int_0^\infty \frac{\gamma(\tau)}{1+j\omega\tau} d\tau, \quad (2)$$

where $R_{pol}$ represents the total polarization resistance and the term $\gamma(\tau)/(1+j\omega\tau)$ represents the relative differential contribution of a single ohmic-capacitive element [29].

In practice, the integral equation in (2) is computed numerically as a finite sum of $n$ RC-elements:

$$Z(j\omega) = R_{ohmic} + R_{pol} \cdot \sum_{k=1}^n \frac{\gamma(\tau_k)}{1+j\omega\tau_k}. \quad (3)$$

The discretization process brings to an intrinsic approximation error that is unavoidable but that can be reduced as much as possible by properly choosing the discretization basis [17], [31]. In [32] Wan et al. studied the effect of discretization methods on DRT calculation, applying piecewise linear functions and radial basis functions (RBFs). They demonstrated that RBFs have more advantages, such as faster convergence and validity overall the time constants domain, including the extension to non-measured points. The main findings of Wan et al. have been exploited in this work to apply RBFs for the problem discretization. The readers are referred to [32] for further details in the mathematical formulation.

Once discretization is applied, the DRT function can be estimated from the impedance measurement by posing a linear inverse problem and solving the following optimization problem:

$$\min_x \|Ax - b\|_2^2, \quad (4)$$

where $\|\cdot\|_2^2$ denotes the Euclidean (L2) norm for vectors, $x$ is an $n$-dimensional vector representing the unknown DRT function, evaluated at the set $\{\tau_k\}_{k=1}^n$ of target time constants, $b$ is an $m$-dimensional vector consisting of the measured impedance points at the frequency set $\{\omega_i\}_{i=1}^m$, and, $A$ is an $m \times n$ matrix whose $A_{ik}$ elements represent the normalized RC elements for the measured frequency $\omega_i$ and the target time constant $\tau_k$ convolved with the RBFs (see [32]).

However, this inverse problem is intrinsically ill-posed [32]. A way to obtain a solution is to apply a numerical regularization. The most used method when applying DRT to Li-ion cells is Tikhonov regularization [29], [30], [32], [33].

The regularization is used to avoid false peaks and oscillations in the solution. The problem in (4) can be rewritten adding a regularization term:

$$\min_x \|Ax - b\|_2^2 + \lambda \cdot x^T M x, \quad (5)$$

where $\lambda$ is the regularization coefficient and $M$ is the regularization matrix. The regularization coefficient $\lambda$ allows to have the best compromise between exactness of the solution and its stability. The choice of this parameter highly influences the final results; in literature three main criteria have been defined to find its optimal value: the real-imaginary part discrepancy test, the real-imaginary cross validation test and the L-curve criterion [34], [35]. The regularization matrix $M$ is used to further improve the problem solution; in the simplest case it is defined as identity matrix. More details on the optimization problem and regularization matrix are given in [34], [36]. For the methodology presented in this work, the L-curve criterion has been chosen for the determination of $\lambda$ and a term proportional to the second derivative of the $\gamma(\tau)$ function has been chosen for the penalty term $x^T M x$.

As stated before, the DRT method works well with ohmic-capacitive electrochemical system. However the inductive behavior at high frequencies and diffusive behavior at low frequencies of Li-ion cells could bring to artefacts in DRT calculation [29]. In both cases, the simplest solution is to truncate the measurement analyzing only the pure ohmic-capacitive part of the spectrum. However, this method leads to lose of valuable information [37]. A second way to proceed is to subtract the inductive contribution from the measurement and to extend the vector of time constants used in DRT calculation. The inductance correction is performed fitting the high frequency part of EIS spectrum with an inductor or with a parallel connection of inductor and resistor [37]. Regarding the diverging tail of the spectrum at low frequency, a used practice is to extend the vector of time constants beyond the minimum and maximum values that corresponds to the minimum and maximum measured frequency points. With this approach there is no extrapolation of EIS points beyond the measured frequency range and the optimization process brings to a stable solution [29].

*B. DRT analysis: ECM configuration*

For a full mapping of the cell behavior, the DRT calculation must be applied to EIS measurements at different SoC (i.e. different OCVs). Once the DRT profiles have been obtained, it is possible to configure the ECM. The configuration is done by analyzing the different peaks of the function $g(\tau)$ that can be associated to the physical processes occurring inside the Li-ion cell. In literature, several works analyzed the time constants range and DRT profiles, allocating four main types of processes [23], [29], [38]–[41]:

- Time constant lower than $10^{-3}$s: particle-particle or particle-current collector interfaces contact resistance. The peculiarity of the peak in this region is its invariance with respect to the SoC of the cell and a small variation due to temperature changes.
- Time constant in the range $10^{-3}$-$10^{-2}$s: the peak represents the transport of Li ions through the SEI layer. There is no a specific variation of the peak linked to the SoC of the cell. However, this peak is sensitive to temperature changes, showing a shift to the right side and an increase in magnitude for low temperatures.
- Time constant in the range $10^{-2}$-$10^{1}$s: charge transfer in the two electrodes: usually two or more peaks are present. In general, the peaks related to the cathode have higher impact (i.e. larger peak) at low SoC, while the ones related to the anode have higher impact at high SoC. At low temperatures charge transfer processes are slowed down due to lower ionic conductivity of the electrolyte and the DRT peaks increase their magnitude. The distinction and position of the peaks within the time constant range depends mainly on the cathode chemistry.
- Time constant higher than $10^{1}$s: diffusive processes. As in the previous case, the peak related to diffusive processes is highly impacted by SoC, depending on the concentration of solid active particles. At low temperatures the diffusivity of ions within the electrodes is reduced and the impedance is higher, implying higher magnitude of the DRT peak.

The allocation of the physical processes and the total number of peaks give a hint on the number of circuit elements that should be adopted in the ECM. For instance, the 5 peaks represented in the DRT profile of Fig. 1 lead to a model including 5 elements. The model can be developed adopting RC branches, given the capacitive-resistive nature of the processes described by the DRT. The main advantages using these simple elements are that the parameters can be calculated from the DRT profile and the direct applicability of the developed model in time-domain simulations. However, in the case of diffusive processes, i.e. at high time constant, the use of a simple RC branch leads to high residuals between the measured and the estimated impedance spectrum. One possible solution to overcome this limitation is to introduce a Warburg element in the model. In the literature, Boukamp et al. demonstrated how the DRT of finite length Warburg element can be derived [42]. This element is represented by a large peak with a high time constant and a series of smaller peaks for decreasing values of time constants that present a defined spacing. This derivation has been exploited in [35], [37] to characterize the peaks related to diffusive processes of the cells under investigation. The authors proved that the time constants of the peaks related to diffusion were compliant with relationship derived in the [42]. In this work a different approach has been chosen. In the frequency domain, the Warburg element with reflective boundary condition can be written as function of the limiting factors $R_{D,R}$ and $C_{D,R}$ [7]:

$$Z_W(j\omega) = \frac{3 \cdot R_{D,R} \cdot \coth\sqrt{3 \cdot j\omega \cdot R_{D,R} \cdot C_{D,R}}}{\sqrt{3 \cdot j\omega \cdot R_{D,R} \cdot C_{D,R}}}. \quad (6)$$

The resistance value $R_{D,R}$ is obtained from the DRT profile, in the region where the time constant is higher than $10^{1}$s. The value of the capacitor $C_{D,R}$ instead, is obtained by differentiation of the OCV curve, as shown in the following equation:

$$C_{D,R}(OCV) = dQ/dV. \quad (7)$$

This parameter, known as the intercalation capacitance, represents the equivalent capacitance of the cell at a specific OCV value. The correspondence between the value of $C_{D,R}$ obtained by OCV differentiation and from EIS measurements at very low frequency have been verified by Brivio et al. in [7]. In this way, the characterization of the cell becomes shorter, avoiding too long impedance measurements or time domain measurements. Another advantage is also on the testing side: it is not necessary to have sophisticated and expensive testing equipment that reach ultra-low frequency values.

Given the objective of developing a time domain model, it is possible to anti-transform the expression in (6) and approximate the Warburg element as a series connection of a capacitor $C_{D,R}$ and an infinite number of RC elements with parameters [43]:

$$C_{n,R} = C_{D,R}/2 \quad and \quad R_{n,R} = 6 \cdot R_{D,R}/(n^2 \cdot \pi^2), \quad (8)$$

where *n* represents the *n-th* element.

The infinite series of RC branches can be limited to a finite number with a negligible effect on the model performances and accuracy [44]. Exploiting the experimental results of [7], in this work we adopted 5 RC branches to describe the Warburg element.

Summarizing all the considerations above, the ECM model for a cell showing *n* peaks in the DRT profile includes:

- A resistor for the ohmic resistance.
- $n-1$ RC branches to model contact resistance, SEI

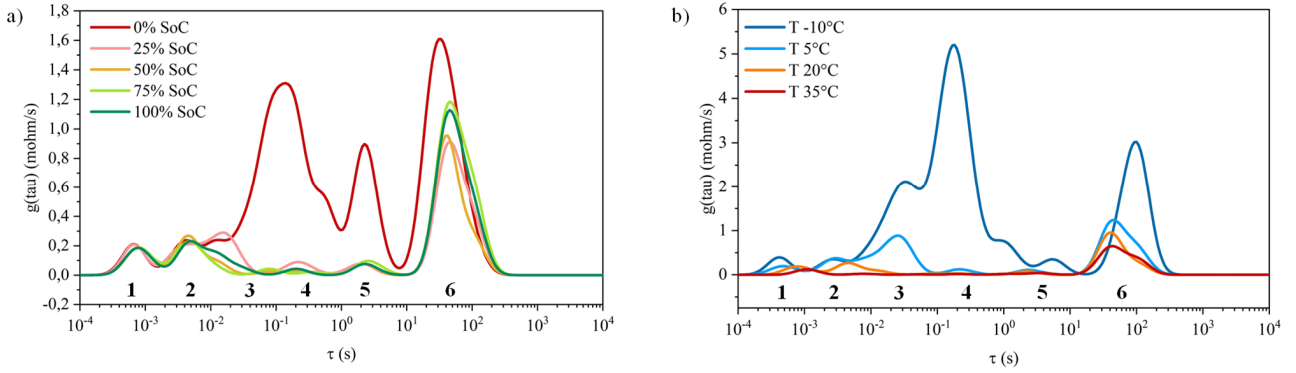

Fig. 2. DRT profiles of the large format pouch NMC cell under investigation. a) DRT profiles at 20°C and different SoC levels; b) DRT profiles at 50% SoC and different SoC levels. At different SoCs and temperatures the cell shows 6 peaks corresponding to: 1) particle-particle, particle-current collector interface; 2) SEI layer; 3-4-5) charge transfer processes; 6) diffusive processes.

layer, charge transfer and double layer effects.
- Series connection of a capacitor and 5 RC elements to approximate the Warburg element exploited to model the diffusion phenomena.

The voltage drop across the capacitor $C_{D,R}$ is used to estimate the OCV value of the model and no voltage generator is needed (i.e. passive electrical model). In turn, the OCV value is used to estimate the SoC of the cell, through the OCV-SoC relationship. This relationship is derived from the OCV curve, computing the SoC as fraction of the whole discharged capacity.

The model can be implemented as a linear system with time-varying parameters that are function of the OCV/SoC [25]. These parameters account for the non-linearities related to the battery dynamic and allow for a simple model implementation. The state space representation of the model is:

$$x(k) = A_k x(k-1) + B_k u(k),$$
$$y(k) = C_k x(k) + D_k u(k), \qquad (9)$$

where the state vector $x(k)$ is defined as:

$$x(k) = \begin{bmatrix} V_{oc}(k) \\ V_1(k) \\ \vdots \\ V_n(k) \end{bmatrix}, \qquad (10)$$

with $V_{oc}$ denoting the OCV and $V_i$ the voltage of the $i$-th RC element. The input $u(k)$ is the current and the observation $y(k)$ is the voltage at the battery terminals at discrete time $k$. The state transition matrix at time $k$ is defined as:

$$A_k = \begin{bmatrix} 1 & 0 & \cdots & 0 \\ 0 & e^{-\frac{T_s}{R_1(k)C_1(k)}} & 0 & 0 \\ \vdots & 0 & \ddots & \vdots \\ 0 & 0 & \cdots & e^{-\frac{T_s}{R_n(k)C_n(k)}} \end{bmatrix} \qquad (11)$$

and the input matrix is defined as:

$$B_k = \begin{bmatrix} \frac{T_s}{C_d(k)} \\ R_1(k)\left[1 - e^{-\frac{T_s}{R_1(k)C_1(k)}}\right] \\ \vdots \\ R_n(k)\left[1 - e^{-\frac{T_s}{R_n(k)C_n(k)}}\right] \end{bmatrix}. \qquad (12)$$

The output matrix $C_k$ is a vector of ones and the feedforward matrix $D_k$ is the ohmic resistance.

### C. Parameters extraction and fitting

The parameters needed for the ECM presented in Section II.B are extracted from the EIS measurement, the DRT profile and the OCV curve.

The ohmic resistance is directly extracted from the EIS curve, looking at the zero-crossing value of the impedance with respect to the real axis in the Nyquist plot. As regard the RC elements, as already mentioned the parameters are extracted by the DRT profile. For the $i^{th}$ peak, the position of its maximum gives the value of the time constant $\tau_i$ and its area gives the value of resistance $R_i$. The value of capacitance is simply computed as:

$$C_i = \tau_i / R_i \qquad (13)$$

A specific algorithm has been developed to improve the estimation of the parameters extracted by the DRT. The algorithm identifies the number of peaks which are present in the DRT profile and performs a fitting with a variable number of gaussian distributions. This number mainly depends on the number of peaks. The main advantage of the algorithm is the ability to distinguish processes also when the peaks are partially overlapped. However, a cross check between the algorithm outcome and the physical processes described in the previous paragraph is always required to ensure that the correct number of peaks is considered.

Once all the parameters are extracted at different SoC values (i.e. OCV values), it is possible to fit them to obtain the parameters trends. An example for $R_i$ and $C_i$ is showed in step C of Fig. 1, respectively with linear and quadratic fitting. The functions obtained are then used by the model during simulations. The only exception is the intercalation capacitance. Its trend is extracted by exploiting (7) as shown in the lower part of step C of Fig. 1. Given the high non-linearity of this parameter, a look-up table is used by the model.

To extend the model validity to different environmental conditions, it is possible to characterize the cell at different temperatures. In this way, the model will include different sets of parameters at given temperatures. When simulating profiles at variable temperature, linear fitting allows to estimate the model parameters over a continuous temperature range between the tested values.

## III. PROPOSED ECM FOR LI-ION NMC CELLS

The methodology presented in Section II has been applied to large format NMC pouch cells [45]. The nominal capacity is 55Ah, the nominal voltage is 3.7V and the voltage range is between 3V and 4.2V.

To develop a generalized model of the battery model under investigation, the characterization tests have been performed on three identical cells. Then, the collected results have been used to calculate the averaged impedance spectra and OCV curves that have been exploited for the model development.

The characterization tests have been performed at the CSEM's energy systems laboratory, located in Neuchâtel, Switzerland, using the following equipment:
- Cell cycler BioLogic BCS815 [46]: battery tester equipped with 32 parallel, 9V-15A channels. ± 0.01% FSD accuracy on the voltage and ± 0.015% FSD accuracy on current, for each available range. EIS capability from 10kHz to 10mHz.
- thermostatic chamber (ATT-DM1200T [47]) with a volume of 1200 L and -45°C/180°C temperature range.

Experiments run at different ambient temperatures: -10°C, 5°C, 20°C and 35°C. A resting phase of 10h has been fixed before the start of the characterization to stabilize the cell temperature. Then, a two steps procedure has been applied:
- Full charge-discharge cycle at C/10 to obtain the OCV curve;
- EIS spectra acquisition at 0%, 25%, 50%, 75% and 100% SoC in the 10mHz-10kHz frequency range (1h rest between the SoC adjustment and the spectrum acquisition).

The OCV protocol has been chosen as an optimal trade-off to reduce the testing time at minimum when compared with longer protocols such as GITT. The OCV curve has been then used to determine the voltage levels corresponding to the chosen SoCs at which the EIS measurements have been run. Looking at test duration, the two steps characterization procedure lasted 44h. The full characterization of the cells at four temperatures has been performed within 10 days.

The data analyses have been performed with dedicated Python scripts, as well as the different steps of the methodology showed in Fig. 1 and described in Section II. All the EIS measurements have been pre-processed to correct the inductive behavior with a parallel circuit of a resistor and an inductor. The L-curve criteria has been applied to define the regularization parameter that has been fixed to $10^{-5}$.

The DRT calculation are shown in Fig. 2: on the left side the DRT profiles at 20°C and different SoC are presented; on the right side the profiles at 50% SoC and different temperature are presented. The numbering under the peaks helps the reader to check the correspondence between the two charts. As stated in Section II, it is possible to link the peaks to the processes occurring in the cell. Peak 1 has negligible variations with respect to the SoC and show a small increase in magnitude decreasing the temperature; it is attributed to particle-particle and particle-current collector contact resistances. All the other peaks instead, show large variations with temperature, with a rise of polarization resistance at low temperatures. This effect is due to the slower reaction rates and diffusion rates. Peak 2 is attributed to the transport of ions through the SEI layer, with small variations varying the SoC. Peak 6 shows large variations at low SoC and it is attributed to the slower diffusion of ions within the electrodes. The remaining three peaks (i.e. number 3,4 and 5) show a large rise at low SoC: this is due to the slower rate of charge transfer processes of the two electrodes. Peak 4 shows the largest variations among the three peaks and at low SoC. For this reason, this peak is attributed to the cathodic charge transfer [23]. More precise information about the attribution of peaks 3 and 5 could be given by measuring EIS spectra of three-electrodes cells or half cells [21], [40].

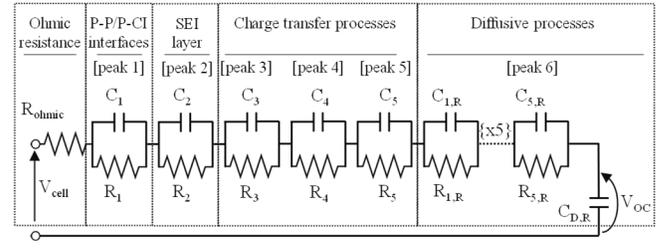

Fig. 4. ECM developed for the large format pouch NMC cell under investigation [43]. The different elements are referred to the modeled physical process and to the related peak of the DRT profiles in Fig. 2.

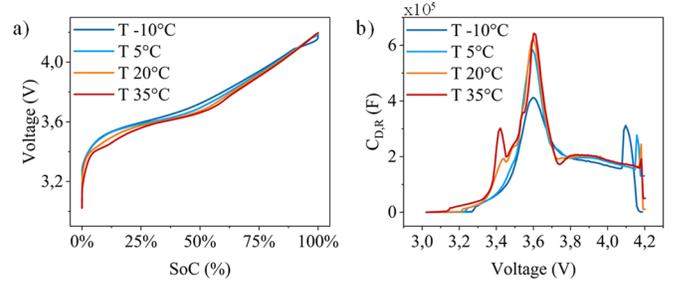

Fig. 3. a) OCV curves of the averaged cell at different temperatures. b) Intercalation capacitance $C_{D,R}$ as function of the voltage obtained from the OCV curves in a) at different temperatures.

Once the peak attribution has been performed, it is possible to proceed with the ECM configuration. The model developed for the cell under investigation is showed in Fig. 4 and it includes 12 elements with a total number of 13 independent parameters:
- A resistor to model the ohmic resistance;
- 1 RC element to represent the particle-particle, particle-current collector interface (peak 1);
- 1 RC element to represent the SEI layer (peak 2);
- 3 RC elements to represent the charge transfer processes (peaks 3, 4, 5);
- 5 RC elements in series with a capacitor ($C_{D,R}$) to approximate the Warburg element representing the diffusive processes (peak 6).

As anticipated in Section II, the voltage drop across the capacitor $C_{D,R}$ is used to estimate the OCV value and in turn the SoC. The OCV of the cell is initialized by the user with a fixed value (the actual SoC of the cell or an estimate of it) and then it is updated at each step of simulation calculating the state vector $x(k)$ defined in (10). The OCV curve (i.e. OCV-SoC relationship) and $C_{D,R}$ profiles at different temperatures are showed in Fig. 3. The SoC has been calculated as fraction of the discharged capacity at C/10 per each different temperature. Analyzing the OCV curves (Fig. 3-a), given a SoC value the corresponding voltage is higher for lower temperatures. This effect is due to the slower cell kinetic at low temperature, and it results in steep voltage variation for SoC lower than 10%. Consequently, the discharged capacity is lower than at ambient conditions. On the contrary, the OCV curve at 35°C presents an additional small plateau region around 10% SoC allowing for a slightly higher value of discharge capacity. The differences between OCV curves reflect also in Fig. 3-b, where the $C_{D,R}$ parameter is investigated. At voltages below 3.4V, the curves at 20°C and 35°C present a peak that is absent in the curves at lower temperatures. At voltages above 4.1V instead, there is a peak with larger magnitude only at lower temperature. Fig. 3-b is also a practical tool to visualize the capacity variation of the due to temperature: the smaller the area under the $C_{D,R}$ curve, the lower the available capacity of the cell.

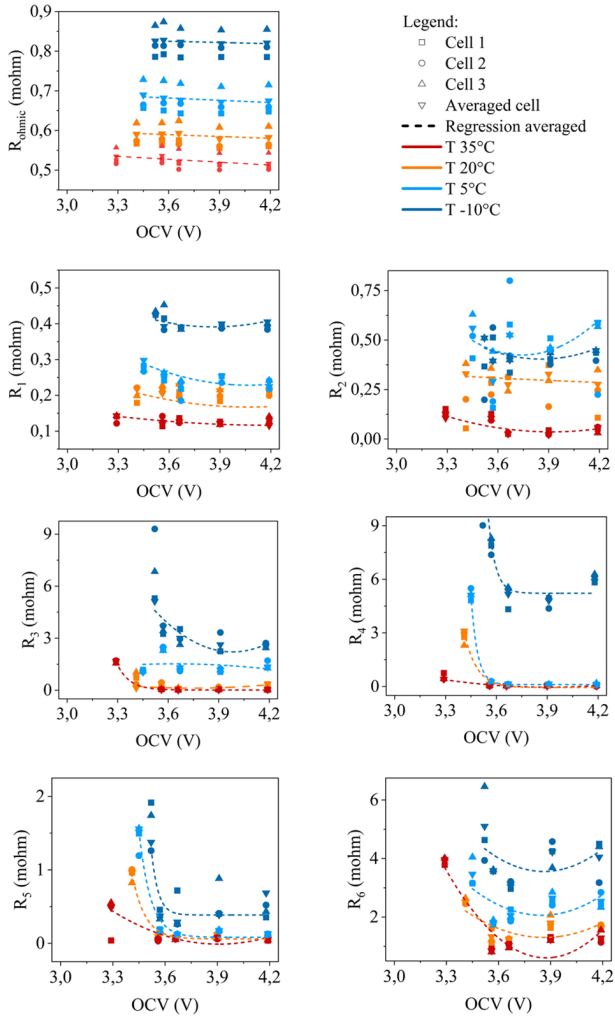

Fig. 5. Resistance parameters extracted by the DRT and ohmic resistance as function of the OCV of the three cells under testing and of the averaged cell representing them. The four temperatures are represented by the different colors. The dashed lines represent the linear or quadratic fitting used to describe the trend.

The ohmic resistance from EIS and the parameters extracted by analyzing the DRT profiles in Fig. 2 are presented in Fig. 5. With lower or higher spread depending on the temperature level, the three tested cells show similar behaviors for all the resistive parameters but slightly different magnitudes. Overall, there is a resistance rise with decreasing temperatures. The ohmic resistance shows a linear trend over the tested OCV range. The spread of measurements around the depicted trend is mainly attributed to the contact between the cell tabs and the holder connected to the cycler. This difference is not affecting all the other parameters given that the ohmic resistance is not part of the $g(\tau)$. The parameter $R_1$ represents the contact resistance and it is characterized by negligible variations over the tested OCV range. $R_2$ is related to the SEI layer and shows well distinct trends at 35°C and 20°C. At lower temperature instead, there is a less clear pattern. The parameters $R_3$, $R_4$ and $R_5$ are related to charge transfer and show a large increase in resistance at low OCV for all temperatures. Moreover, $R_3$ and $R_4$ show similar values at mid- and high temperatures and much higher values at low and very low temperatures. As regards the diffusive processes, $R_6$ shows similar quadratic behavior at different temperatures, with higher magnitude at low and high voltages and a minimum around 3.9V.

The parameters' trends shown in Fig. 5 have been given as inputs for the model simulations.

As mentioned in Section II-C, when dealing with variable temperature simulations, the parameters' estimation over a continuous temperature range is performed with linear fitting between the two closest characterized values.

## IV. VALIDATION RESULTS

The model developed in Section II has been validated with time domain simulations. Three types of current profiles have been tested: dynamic profile (i.e. current steps), full discharge and driving profile. These profiles have been applied at the four temperature levels chosen for cell characterization (i.e. -10°C, 5°C, 20°C and 35°C). Moreover, the driving profile has been also applied under variable temperature conditions to simulate different seasons. The details about the testing protocols and the tests results will be given in the next sub-sections.

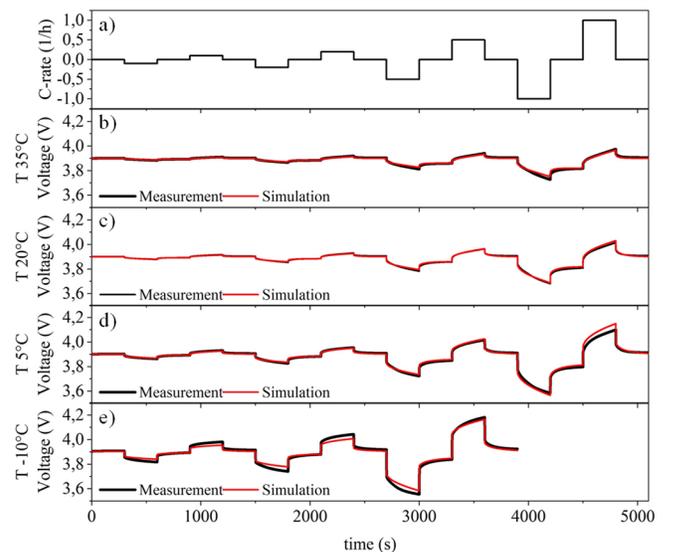

Fig. 6. Dynamic test at four C-rates (C/10, C/5, C/2 and 1C) with 5 minutes steps of charge and discharge spaced by resting phase. a) Current profile. Measured and simulated voltage profiles at: b) 35°C; c) 20°C; d) 5°C and e) -10°C.

TABLE I
AVERAGED RMSE ERRORS OF THE DYNAMIC AND DRIVING PROFILES AT DIFFERENT SoCs AND OF THE FULL DISCHARGE AT 0.1C. THE RMSE VALUES ARE GIVEN AS ABSOLUTE VALUE (MV) AND AS PERCENTAGE OF THE VOLTAGE INTERVAL EXPLOITED BY THE CELL (3-4.2V) AT THE FOUR TESTED TEMPERATURES. AT -10°C THE DYNAMIC PROFILES INCLUDE ONLY 3 CURRENT STEPS: C/10, C/5 AND C/2 (DEPICTED WITH *).

| Temperature | RMSE | Dynamic profile | | | Driving profile | | | Full discharge |
|---|---|---|---|---|---|---|---|---|
| | | 75% SoC | 50% SoC | 25% SoC | 75% SoC | 50% SoC | 25% SoC | |
| 35°C | absolute (mV) | 6.46 | 6.07 | 14.35 | 2.70 | 1.47 | 3.64 | 9.84 |
| | % of voltage interval | 0.54% | 0.51% | 1.20% | 0.22% | 0.12% | 0.30% | 0.82% |
| 20°C | absolute (mV) | 5.99 | 8.86 | 16.56 | 2.72 | 2.44 | 6.03 | 17.54 |
| | % of voltage interval | 0.50% | 0.74% | 1.38% | 0.23% | 0.20% | 0.50% | 1.46% |
| 5°C | absolute (mV) | 12.93 | 20.26 | 35.05 | 5.20 | 7.00 | 22.56 | 38.23 |
| | % of voltage interval | 1.08% | 1.69% | 2.92% | 0.43% | 0.58% | 1.88% | 3.19% |
| -10°C | absolute (mV) | 19.83* | 41.07* | 64.26* | 32.26 | 41.31 | 19.18 | 35.21 |
| | % of voltage interval | 1.65% | 3.42% | 5.35% | 2.69% | 3.44% | 1.60% | 2.93% |

### A. Dynamic profile

The dynamic testing profile has been performed with a repetition of 4 steps sequence including rest, discharging, rest and charging phases. The duration of each step has been fixed to 5 minutes. Four C-rates have been applied: C/10, C/5, C/2 and 1C. The obtained current profile is showed in Fig. 6-a. The measured and simulated voltage profiles at 35°C, 20°C, 5°C and -10°C are showed respectively in Fig. 6-b, c, d and e. The profile at -10°C has been performed only up to C/2 as indicated in the cell datasheet [45]. In general, the model is able to simulate the cell dynamics in the full range of currents and temperatures. It shows a good level of accuracy at 20°C and 35°C: the absolute RMSE values are lower than 10mV at 50% and 75% SoC (TABLE I). At lower SoC, the maximum error is 17 mV at 20°C. Considering lower temperatures, additional non-linearities occur in the cell behavior due to slower dynamic, impacting the model accuracy. At -10°C and 25% SoC, the dynamic profile simulation showed 64.26mV of absolute error (5.35% with respect to the voltage interval). This is the maximum error registered over the 12 run simulations.

### B. Full discharge profile

The full discharge tests have been performed at the four defined temperatures with a constant rate of C/10. In addition, higher current rates have been analyzed at 20°C: C/5, C/2 and 1C. Looking at the results at C/10, the measured and simulated voltage profiles are shown in Fig. 8. The main difference between the four temperatures is the duration of the test: the lower the temperature, the lower the test duration due to less available capacity. The model well approximates the cell behavior over the whole voltage range, with the highest discordances at low voltage. The absolute RMSE shows its minimum at 35°C (9.84mV) and its maximum at 5°C (38.23mV).

The validation results at 20°C and different rates are compared in Fig. 7. The model shows good performances following the voltage trend of the real cell. The RMSE values are listed in TABLE II: the error is lower than 3% (i.e. 36mV) up to C/2 and shows its maximum at 1C (61.68mV).

### C. Driving profile

The driving profile has been derived combining the dynamic discharge profile A and B for BEV specified in the IEC 62660-

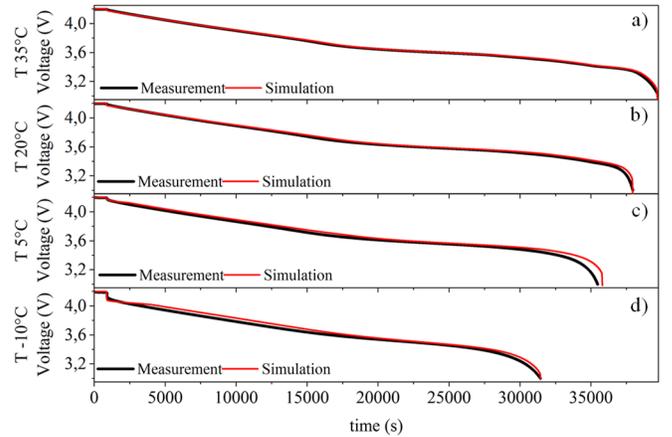

Fig. 8. Full discharge tests at constant current rate of C/10. Measured and simulated voltage profiles at: a) 35°C; b) 20°C; c) 5°C and d) -10°C.

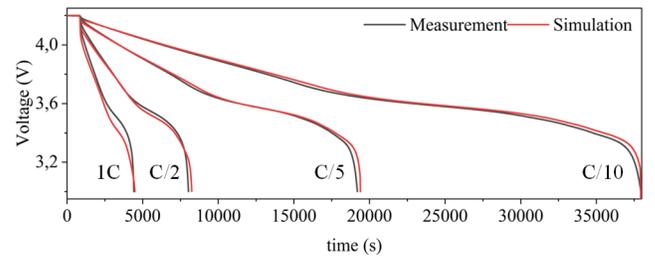

Fig. 7. Measured and simulated voltage profiles of full discharge tests at 20°C and different current rates: 1C, C/2, C/5 and C/10.

TABLE II
AVERAGED RMSE ERRORS OF THE DISCHARGE PROFILE AT 20°C AND DIFFERENT CURRENT RATES

| RMSE | Current rate | | | |
|---|---|---|---|---|
| | C/10 | C/5 | C/2 | 1C |
| absolute (mV) | 17.54 | 28.07 | 34.62 | 61.68 |
| % of voltage interval | 1.46% | 2.34% | 2.88% | 5.14% |

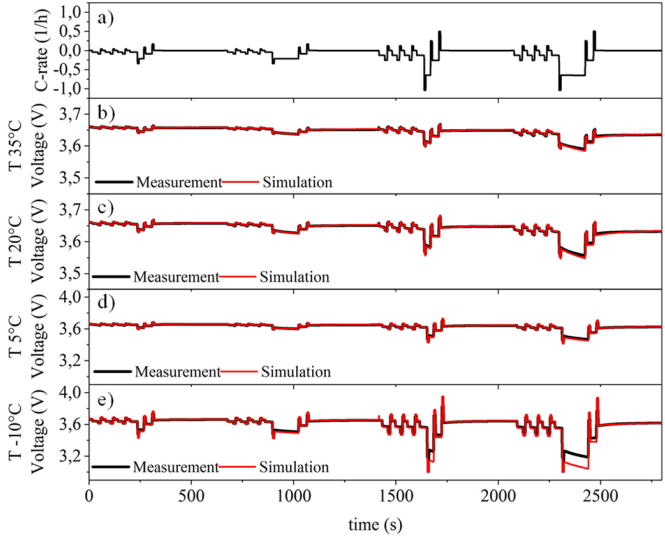

Fig. 9. Driving simulation test. a) Current profile derived by IEC 62-660-1 regulation with repetition of profile A and B at different maximum rates. Measured and simulated voltage profiles at: b) 35°C; c) 20°C; d) 5°C and e) -10°C.

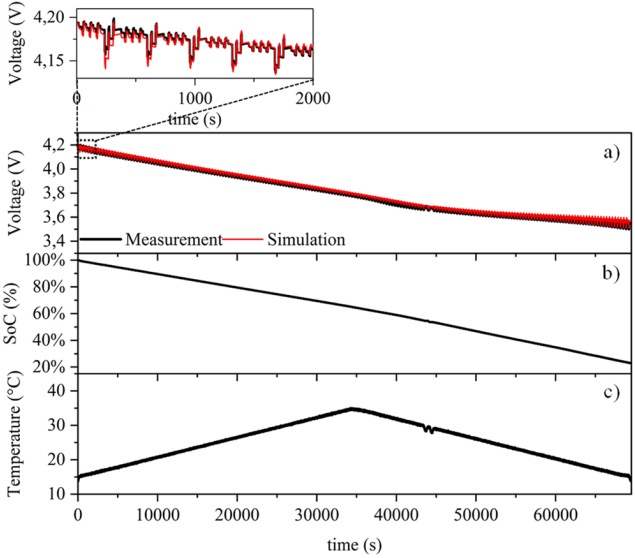

Fig. 10. Driving simulation test at variable temperature. a) Measured and simulated voltage profiles; b) SoC estimated by the model; c) temperature profile measured on the cell surface in the summer condition simulation (temperature between 15°C and 35°C).

TABLE III
AVERAGED RMSE ERRORS OF THE DRIVING SIMULATION PROFILES PERFORMED AT VARIABLE TEMPERATURE

| RMSE | T range [15°C,35°C] (summer) | T range [5°C,25°C] (spring/fall) | T range [-5°C,15°C] (winter) |
|---|---|---|---|
| absolute (mV) | 11.18 | 18.01 | 14.47 |
| % of voltage interval | 0.93% | 1.50% | 1.21% |

In the second case, i.e. at variable temperature, the testing protocol has been developed as follows:
- Repetition of Profile A (*N=1/3*) until the cell reaches 50% SoC;
- Resting phase 5 minutes;
- Profile B (*N=1/3*)
- Repetition of Profile A (*N=1/3*) until the cell reaches 20% SoC.

The temperature profiles have been developed with symmetrical heating and cooling ramps (2°C/h) and repeated over three different temperature ranges to simulate summer (15°C-35°C), spring/fall (5°C-25°C) and winter (-5°C-15°C) scenarios. RMSE results are given in TABLE III. The model shows high accuracy in all the three different conditions, with a maximum error of 18mV, corresponding to 1.5% of the voltage range. The measured and simulated voltage profiles are given in Fig. 10-a. Moreover, given the long duration of the simulation also the SoC estimation is reported for this case (Fig. 10-b). SoC is directly derived from the model by exploiting the SoC-OCV relationship.

## V.  CONCLUSIONS

This work introduced EIS2MOD, a modelling framework to develop a physically based model from EIS and OCV measurements for Li-ion cells. DRT has been applied to deconvolve EIS spectra as an efficient technique to distinguish electrochemical phenomena which are difficult to interpret directly from EIS.

The analysis of DRT deconvolutions at different SoCs and temperatures allowed to define an ECM in which each circuit element is assigned to a physical phenomenon. Moreover, the model parameter determination is a direct consequence of the DRT post-processing, thus avoiding any fitting process of the EIS and simplifying the model implementation for simulation purposes. The model developed for large format NMC pouch cells constitutes of five blocks to represent the different phenomena with a total number of 13 independent parameters. These parameters are function of the OCV (i.e. SoC) and the temperature. The model validation has been performed with different current profiles at constant and variable temperature. In the first case, simulations at 20°C and 35°C showed RMSE from 0.2% to 1.5% for dynamic profiles (at different SoCs) and maximum RMSE of 5.1% for full discharge at 1C. At lower temperatures the high non-linearities in battery dynamic influenced the model behavior and a maximum RMSE of 5.4% was obtained for dynamic simulation at -10°C. In the second case, driving profile simulations at variable temperature showed a maximum RMSE of 1.5%.

Future works will concentrate on two main directions. On one side, a first objective will be to extend the methodology for SoH estimation by exploiting DRT analysis. Preliminary results

1 standards about performance testing of Li-ion cells for propulsion of electric vehicles [26]. The maximum power output defined in these two profiles depends on the parameter *N*, that represents the fraction of capacity that can be used with respect to the maximum capacity of the cell. Two groups of experiments have been performed to validate the cell model at constant and at variable temperature.

In the first case, the testing protocol has been developed with series repetition of profile A and B assuming *N=1/3* and *N=1*. The current and voltage profiles at different temperatures are shown in Fig. 9. As in the previous cases the model shows higher accuracy at 20°C and 35°C than at 5°C and -10°C. When considering the simulation at SoC higher than 25% and temperature between 5°C and 35°C, the maximum RMSE is 7mV (less than 0.6% of the voltage range).

of cycling aging tests show that DRT profiles well represent the degradation phenomena with shifts and changes in magnitude of the peaks. These variations will be used to adjust the model parameters over lifetime and to allow not only for an accurate SoC estimation, but also SoH estimation. On the other side, a second objective will be to port the code in a dedicated BMS hardware/software solution that can leverage its potentiality by implementing additional sensing capabilities at cell level.